\documentclass[draftcls, onecolumn, 10pt]{IEEEtran}
\usepackage{amssymb,amsmath}
\usepackage[dvips]{graphicx}
\usepackage{amsfonts}
\usepackage{subfigure}
\usepackage{booktabs,multirow}
\usepackage{color}
\usepackage{cite}
\usepackage[table]{xcolor}

\IEEEoverridecommandlockouts

\begin{document}

\title{How Much of Wireless Rate Can Smartphones Support in 5G Networks?}

\author{\normalsize
Jing Yang$^1$,~\IEEEmembership{Student~Member,~IEEE,} Xiaohu Ge$^1$,~\IEEEmembership{Senior~Member,~IEEE,} Yi Zhong$^1$,~\IEEEmembership{Member,~IEEE}\\
\vspace{0.70cm}
\small{
$^1$School of Electronic Information and Communications,\\
Huazhong University of Science and Technology, Wuhan 430074, Hubei, P. R. China.\\
%Email: \{xhge \}@mail.hust.edu.cn\\
Email: \{yang\_jing, xhge, yzhong\}@mail.hust.edu.cn}\\
%\vspace{0.2cm}
%$^2$National Institute of Standards and Technology
%(NIST),\\
%Gaithersburg, MD 20899-8920 USA.\\
%Email: hamid.gharavi@nist.gov}\\
%\vspace{0.2cm}
%$^3$Institute of Sensors, Signals and Systems, \\
%School of Engineering \& Physical Sciences,
%Heriot-Watt University, Edinburgh, EH14 4AS, UK.\\
%Email: cheng-xiang.wang@hw.ac.uk}\\
\thanks{\small{Submitted to IEEE Network.}}
\thanks{\small{Correspondence author: Dr. Xiaohu Ge, Email: xhge@mail.hust.edu.cn }}
}

%\date{\today}
\renewcommand{\baselinestretch}{1.2}
\thispagestyle{empty}
\maketitle
\thispagestyle{empty}
%\newpage
\setcounter{page}{1}\begin{abstract}
Due to the higher wireless transmission rates in the fifth generation (5G) cellular networks, higher computation overhead is incurred in smartphones, which can cause the wireless transmission rates to be limited by the computation capability of wireless terminals. In this case, is there a maximum receiving rate for smartphones to maintain stable wireless communications in 5G cellular networks? The main objective of this article is to investigate the maximum receiving rate of smartphones and its influence on 5G cellular networks. Based on Landauer's principle and the safe temperature bound on the smartphone surface, a maximum receiving rate of the smartphone is proposed for 5G cellular networks. Moreover, the impact of the maximum receiving rate of smartphones on the link adaptive transmission schemes has been investigated. Numerical analyses imply that the maximum receiving rate of smartphones cannot always catch up with the downlink rates of future 5G cellular networks. Therefore, the link adaptive transmission scheme for future 5G cellular networks has to take the maximum receiving rate of smartphones into account.
\end{abstract}

\IEEEpeerreviewmaketitle

\newpage
\section{Introduction}
Currently, the peak rate of the fifth generation (5G) communication systems is expected to reach 20 gigabits per second (Gbps) \cite{1Ge}. Moreover, the future beyond 5G wireless communications is expected to reach 100 Gbps \cite{2Koenig} for short-range communication by utilizing the terahertz (THz) bands. Based on high transmission rates, some high-data-rate applications, such as virtual reality (VR) and augmented reality (AR), will be applied to wireless terminals, e.g., smartphones \cite{3Zhong}. It is inevitable that smartphones will confront a dramatic data receiving and processing challenge in 5G and future 6G cellular networks. Moreover, the computation capability of smartphones is limited, making it a great challenge for smartphones to process massive wireless data in 5G cellular networks \cite{4Ge}. Therefore, new issues and limits triggered by smartphones are emerging for 5G cellular networks.

Current and previous studies related to the capacities of cellular networks have a default assumption that the maximum receiving rate of smartphones always catches up with the downlink rate of base stations (BSs), irrespective of the downlink rate. In general, the maximum receiving rate of smartphones is the rate at which the data can be stably processed by chips in smartphones. Therefore, two factors restrict the maximum receiving rate of smartphones. One factor is the computation capability of the baseband processor, and the other is heat dissipation.

The chip in smartphones has integrated all components of computations and communications, such as the application processor (AP), storage unit and baseband processor (BP). When wireless communications run on smartphones, some critical processes related to receiving rates, such as digital signal processing and channel coding processes, are carried out by the BP. Thus, the maximum receiving rate of smartphones is limited by the computation capability of the BP. Moreover, the computation capability of the BP depends on the semiconductor technology. On the basis of utilizing the latest semiconductor technology to produce microchips, more transistors are being integrated into a microchip to increase the computation capability. Although the evolution of semiconductor technology has followed Moore's law, the death of Moore's law has been announced by the verification of the Landauer limit \cite{5Berut} and the influence of thermal noise \cite{6Izydorczyk}. Additionally, silicon transistors will approach a projected scaling limit of 5-nanometer (nm) gate lengths \cite{7Desai,8Qiu}. It is predicted that a chip based on 5-nm semiconductor technology will be produced around 2020, which will launch 5G cellular networks. Thus, the computation capability of the BP in smartphones will approach the limit for future 5G cellular networks.

In addition to the computation capability limit, the maximum receiving rate is restricted by the heat dissipation of smartphones. Since wireless terminals have entered into the smart era, the development of smartphones is accompanied by overheating issues. In general, the major heat contributing to the overheating issues in smartphones is caused by chip computation. The relationship between heat generation and computation can be interpreted clearly by Landauer's principle. Furthermore, the temperature of the smartphone surface has been limited to 45 $^\circ {\rm{C}}$ \cite{9Chiriac}. When the temperature of a smartphone's surface surpasses 45 $^\circ {\rm{C}}$, it is necessary to diminish the heat generated by the chip. Consequently, smartphones must reduce the computation capability of the chip to decrease the heat generation. In some extreme conditions, the receiving rates of smartphones are cut down or stopped by a protection scheme. Therefore, the heat dissipation of smartphones restricts the maximum receiving rate of smartphones. Although the maximum receiving rate of smartphones is restricted by the computation capability and heat dissipation, detailed studies of basic models used for evaluating the maximum receiving rate of smartphones are surprisingly rare in the available literature. Moreover, the impact of the computation capability and heat dissipation of smartphones on 5G cellular networks has not been investigated.

Inspired by this gap in knowledge, we first analyze the computation capability and heat dissipation of smartphones in detail. Then, a maximum receiving rate is derived for sustaining stable computations and communications at the smartphones. Moreover, the impact of the maximum receiving rate of smartphones on the link adaptive transmission scheme is analyzed. Finally, future challenges with respect to the maximum receiving rate of smartphones are discussed, and conclusions are drawn in the last section.

\section{Computation Capability Factor}
Before estimating the maximum receiving rate of smartphones, the computation capability of the BP needs to be analyzed. In this section, Landauer's principle and the semiconductor technology limits will be incorporated to analyze the computation capability of the BP.

\subsection{Computation Capability Analysis}

\begin{figure}[!t]
\begin{center}
\includegraphics[width=5.5in]{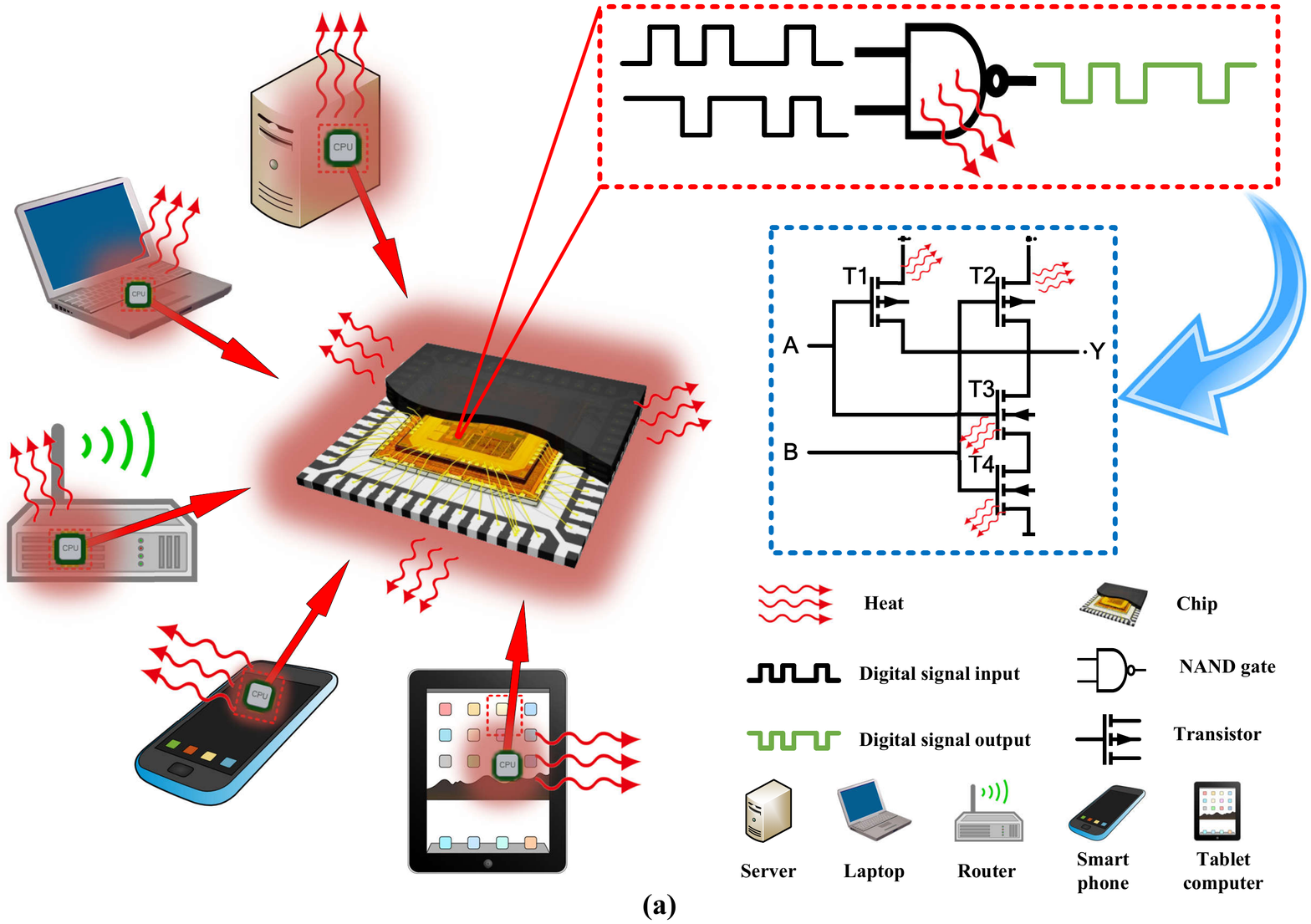}
\includegraphics[width=5.5in]{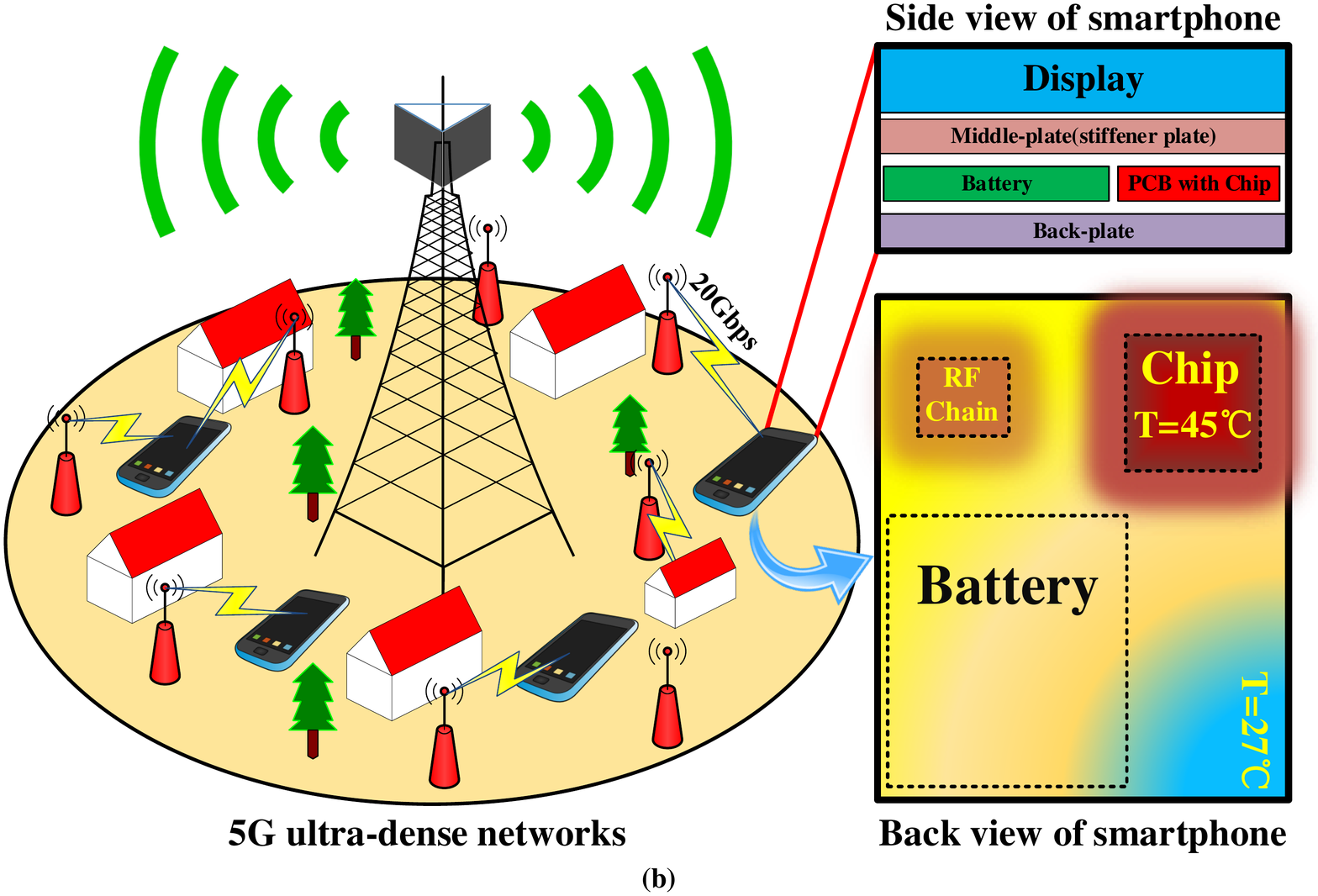}
\caption{(a) Computation and heat generation of electronic products; (b) Structure and temperature distribution of smartphones.}\label{Fig1}
\end{center}
\end{figure}

The biggest heat generated by electronic products, such as laptops and smartphones, comes from the computation of the chips (Fig. 1a). When the 5G base stations communicate with smartphones at the peak rate, the chip in smartphones has to confront dramatic computations that can generate a large amount of heat and raise the temperature of the chip (Fig. 1b). For the same semiconductor technology, the computation capability of different chips can be reflected by the heat generation rate. Moreover, the relationship between the computation capability and heat generation can be established by Landauer's principle.

Landauer's principle, proposed in 1961 by Rolf Landauer \cite{10Ge}, is a physical principle that indicates that any logically irreversible manipulations of information, such as the erasure of a bit or merging of two computation paths, must be accompanied by a corresponding entropy increase in the environment \cite{11Bennett}. Based on thermodynamics, the entropy increase in the environment is in the form of heat. Moreover, the energy used to erase one bit of information has a lower bound, which is known as the Landauer limit: $kT\ln (2)$, where $k$ is the Boltzmann constant, that is, $1.38 \times {10^{ - 23}}\;{\rm{J}}/{\rm{K}}$, and $T$ is the temperature in Kelvin. Furthermore, the Landauer limit has been proved as the lower bound of the transistor switching energy \cite{5Berut}. The switching energy of the transistors produced with the latest semiconductor technology still has a gap of two orders of magnitude to approach the Landauer limit. In this article, the gap between the transistor switching energy and Landauer limit is denoted ${G_{\rm{S}}}$, where the subscript $S$ represents the level of semiconductor technology. For instance, ${G_{10}}$ indicates the gap between the 10-nm semiconductor technology and Landauer limit.

Based on Landauer's principle and $S$ nm semiconductor technology, more computations cause more heat generation at the chip, which consumes more computation power. Accordingly, the computation capability can be indicated by the computation power consumption of the chip $P_{{\rm{comp}}}^{{\rm{chip}}}$, which mainly consists of the power consumption of the AP ${P_{{\rm{AP}}}}$, storage unit ${P_{{\rm{Storage}}}}$ and BP ${P_{{\rm{BP}}}}$. Among these three parts, the maximum receiving rate of smartphones is mainly limited by the computation capability of the BP. Before estimating the ${P_{{\rm{BP}}}}$ by Landauer's principle, it is necessary to consider the sum of computed or erased information in baseband processor ${C_{{\rm{BP}}}}$, ${C_{{\rm{BP}}}} = {K_{{\rm{BP}}}}{R_{{\rm{phone}}}}$, where ${R_{{\rm{phone}}}}$ is the receiving rate of smartphones and ${K_{{\rm{BP}}}}$ is the logic operations per bit of the algorithm in the BP that can be achieved, approximately ${10^8}$ \cite{12Mammela}. Therefore, the computation power consumption of the BP can be calculated by ${P_{{\rm{BP}}}} = {C_{{\rm{BP}}}}{F_0}\alpha {G_{\rm{S}}}kT\ln 2$, where ${F_0}$ is the number of loading logic gates, known as the fanout, whose typical value is 3-4, and $\alpha $ is the activity factor whose typical range is 0.1-0.2 \cite{12Mammela}.

\subsection{Semiconductor Technology Limits}
The evolution of semiconductor technology has promoted the development of smart terminals, and Moore's law has affected the semiconductor technology in the last fifty years. However, Moore's law was invalidated by the verification of the Landauer limit and the effect of thermal noise \cite{6Izydorczyk,12Mammela}. The Landauer limit shows that the switching energy of transistors has a lower bound \cite{12Mammela}. Moreover, the thermal noise death of Moore's law was proved from the viewpoint of communication in terms of transmitting the internal signals of chips correctly with non-negligible effects of thermal noise \cite{6Izydorczyk}.

As shown in Fig. 2, Moore's law started to fail around 2015, and the silicon transistors approach the projected scaling limit of 5-nm gate lengths \cite{7Desai,8Qiu}. Moreover, silicon transistors, with the limit of 5-nm gate lengths, still have a gap of two orders of magnitude between the transistor switching energy and the Landauer limit. This gap can be estimated as ${G_5} \approx 454.2$ \cite{8Qiu}. Although 1-nanometer gate length transistors have been invented with molybdenum disulfide (${\rm{Mo}}{{\rm{S}}_2}$) instead of silicon \cite{7Desai}, applying the raw material ${\rm{Mo}}{{\rm{S}}_2}$ to chips will take a long time because it is difficult to integrate billions of transistors with new material into the chip. Therefore, the gap ${G_{\rm{S}}}$ cannot be resolved until a long time after silicon transistors reach the projected scaling limit.

When the gap ${G_{\rm{S}}}$ cannot be diminished, the computation power consumption of the chip in smartphones increases with the growth of the data rates in 5G wireless communication systems. The peak rate is expected to reach 20 Gbps in 5G cellular networks and exceeds 100 Gbps for subsequent networks (Fig. 2). The higher wireless transmission rate implies that the computation power consumption will be increased in both smartphones and BSs. When the downlink rate is less than the Gbps level, the computation power consumption of chips is commonly treated as a small circuit power consumption or ignored. When the downlink rate in 5G cellular networks achieves 20 Gbps, the computation power consumption of the chips cannot be ignored or fixed as a constant.

\begin{figure}[!t]
\begin{center}
\includegraphics[width=6.5in]{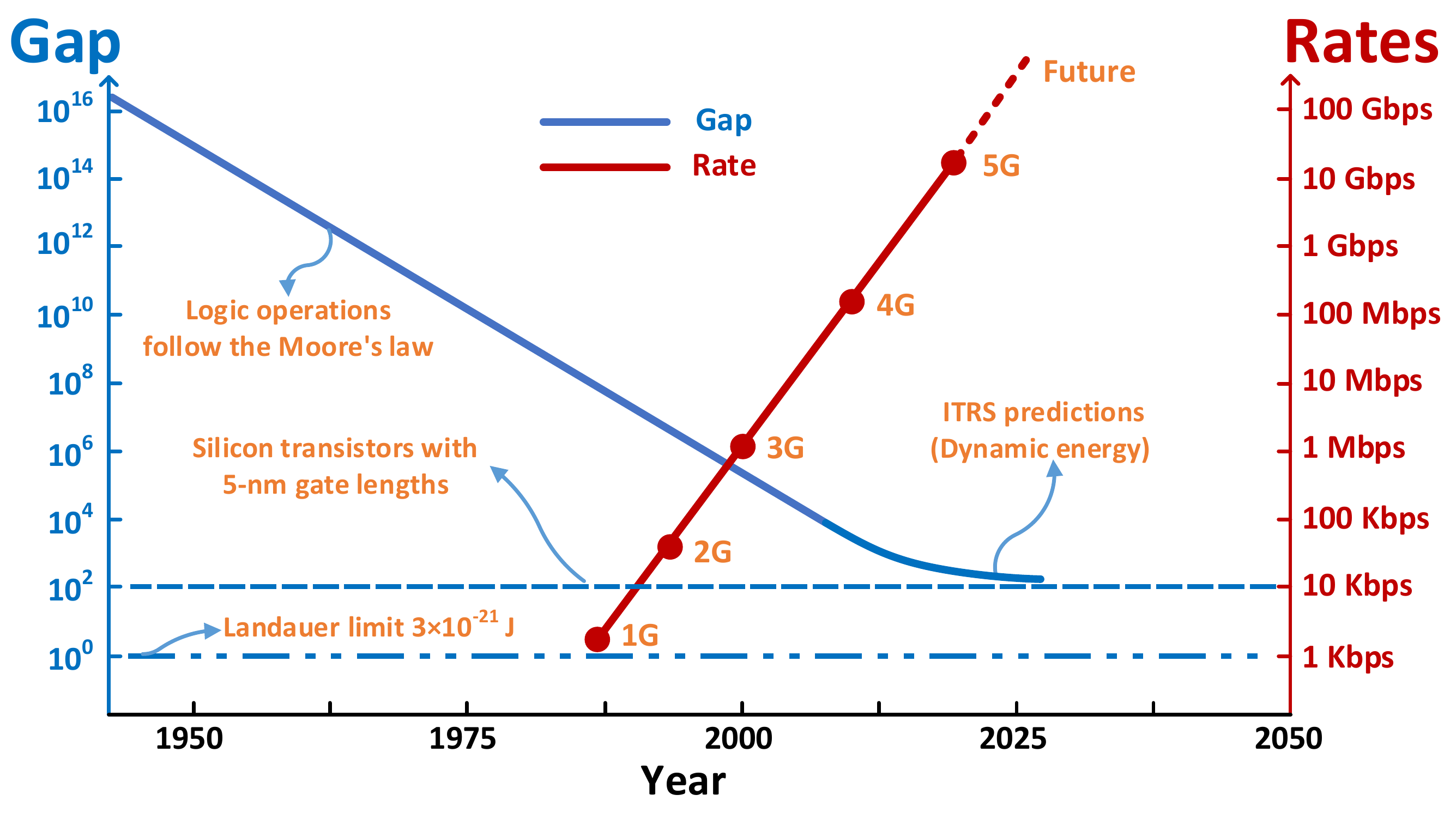}
\caption{The gap between the transistor switching energy and Landauer limit VS. wireless transmission rate.}\label{Fig2}
\end{center}
\end{figure}

\section{Heat Dissipation Factor}
Analyzing the heat generation and transfer processes in smartphones is necessary to assess the restriction of heat dissipation on the maximum receiving rate of smartphones. In this section, the heat generation and transfer processes with respect to radio frequency (RF) chains and smartphone chips have been investigated.

\subsection{Heat Generation and Transfer on RF Chains}
In the downlink, smartphones passively receive the wireless signals from the base station. The received wireless signals cannot be directly processed by smartphones due to channel fading. Consequently, the smartphones utilize RF chains, including low noise amplifiers (LNAs), to amplify the received wireless signals. In this case, the major heat generated by RF chains comes from the LNAs. Without loss of generality, each antenna is configured with an RF chain at the smartphone. Therefore, the heat generation of the LNAs can be calculated by ${H_{{\rm{LNA}}}}{\rm{ = }}N_{{\rm{TRX}}}^{{\rm{phone}}}{P_{{\rm{LNA}}}}\left( {1 - \eta } \right)$, where $N_{{\rm{TRX}}}^{{\rm{phone}}}$ is the number of antennas, ${P_{{\rm{LNA}}}}$ is the power of an LNA and $\eta $ is the power-added efficiency of the LNAs.

Considering the electromagnetic compatibility in smartphones, RF chains are usually separated from the chip of smartphones. Hence, the ratio of the heat transferred from the LNAs to the chip can be configured as $\lambda $ in this paper. Based on Fourier's law in the heat transfer theory, the value of $\lambda $ can be estimated using the thermal conductivity of the printed circuit board (PCB) and the heat conduction distance between the LNAs and chip.

\subsection{Heat Generation and Transfer on Chips}
Based on Landauer's principle, irreversible computations increase the entropy in the environment by generating heat. There is no other heat generation, such as friction of mechanical motion, during the computation of chips. Therefore, the heat generation power of the chip $H_{{\rm{Thermal}}}^{{\rm{chip}}}$ is equal to the computation power consumption of the chip $P_{{\rm{comp}}}^{{\rm{chip}}}$.

To sustain stable computations and communications in smartphones, heat dissipation is required. When the heat has been generated by the computations of the chip, the heatsink in smartphones transfers the heat from the chip to the smartphone surface and other low-temperature components, e.g., the battery and PCB, with heat conduction rates $Q_{{\rm{conduction}}}^{{\rm{sur}}}$ and $Q_{{\rm{conduction}}}^{{\rm{components}}}$, respectively. Moreover, the sum of $Q_{{\rm{conduction}}}^{{\rm{sur}}}$ and $Q_{{\rm{conduction}}}^{{\rm{components}}}$ can be approximated by the value of $H_{{\rm{Thermal}}}^{{\rm{chip}}}$. Limited by the thickness of smartphones, the heat conduction distance from the chip to the surface of smartphones is only a few millimeters, which is shorter than the distance from the chip to other components. In this case, the value of $Q_{{\rm{conduction}}}^{{\rm{sur}}}$ is larger than the value of $Q_{{\rm{conduction}}}^{{\rm{components}}}$ according to the heat transfer theory. Hence, most of the heat is transferred from the chip to the smartphone surface by the heatsink. Then, the heat accumulated on the smartphone surface is removed by free air convection at a rate ${Q_{{\rm{convection}}}}$. The size of smartphones is too small to use active cooling, such as forced convection caused by a fan, to dissipate the heat from smartphones. Therefore, free air convection, known as passive cooling technology, is the only technique to dissipate heat out of smartphones.

Considering that the chip is the highest temperature component in smartphones (Fig. 1b), the thermal design of smartphones mainly focuses on heat dissipation from the chip. Hence, the maximum heat dissipation power of the chip is configured as the thermal design power of smartphones ${P_{{\rm{TD}}}}$, which is equal to the sum of $Q_{{\rm{conduction}}}^{{\rm{components}}}$ and ${Q_{{\rm{convection}}}}$. When the value of $H_{{\rm{Thermal}}}^{{\rm{chip}}}$ is larger than the value of ${P_{{\rm{TD}}}}$, heat accumulates on the surface of smartphones, and the temperature of the smartphone surface increases. To prevent a local high-temperature area on the smartphone surface, the thermal design of smartphones must turn smartphones into an isothermal body. However, a local high-temperature area on the smartphone surface always occurs near the chip (Fig. 1b). When the temperature anywhere on the smartphone surface exceeds 45 $^\circ {\rm{C}}$, a low-temperature burn is likely to occur on the users' skin that touches the local high-temperature area on the smartphone surface \cite{13Moritz}. Thus, the highest temperature of the smartphone surface is limited to 45 $^\circ {\rm{C}}$ \cite{9Chiriac}, which can be regarded as a safe temperature bound ${T_{{\rm{safe}}}}$.

Restricted by the passive cooling and safe temperature bound on the smartphone surface, the thermal design power of smartphones is a challenge to optimize, especially considering the characteristics of the chip in smartphones, e.g., small size and high heat density. Although the power consumption of the chip has been scaled down by utilizing the latest semiconductor technology, the heat density of chips still has not been effectively reduced. The parameters of different chip products from various companies are illustrated in Table I. The heat density of a full-load Qualcomm Snapdragon 835 used in smartphones is larger than that of the Intel\textregistered{} Core\texttrademark{} i7-7920HQ used in laptop computers. When smartphones have a high receiving rate, a large amount of heat is generated by the chip on a 1 ${\rm{c}}{{\rm{m}}^2}$ surface area. Hence, $H_{{\rm{Thermal}}}^{{\rm{chip}}}$ has a large probability of exceeding ${P_{{\rm{TD}}}}$.

\begin{table}
\begin{centering}
\begin{tabular}{|c|c|c|c|c|c|c|}
\hline
 &  &  & Semiconductor &  & Package & Heat\tabularnewline
Devices & Companies & Chip Products & Technology & Power & Size & Density\tabularnewline
 &  &  & (nm) & (W) & ($\rm{c\ensuremath{m^{2}}}$) & ($\rm{W/c\ensuremath{m^{2}}}$)\tabularnewline
\hline
Server  &  & Xeon\textregistered{} Processor E7-8894 v4  & 14 & 165 & 23.63 & 6.89\tabularnewline
\cline{1-1} \cline{3-7}
Laptop  & Intel & Core\texttrademark{} i7-7920HQ  & 14 & 45 & 11.76 & 3.83\tabularnewline
\cline{1-1} \cline{3-7}
Tablet  &  & Core\texttrademark{} m3-7Y32 Processor  & 14 & 4.5 & 3.30 & 1.36\tabularnewline
\hline
 & Qualcomm & Snapdragon 835  & 10 & 3.6 & 0.72 & 5.00\tabularnewline
\cline{3-7}
 &  & Snapdragon 820  & 14 & 4.6 & 1.14 & 4.04\tabularnewline
\cline{2-7}
 &  & A10 Fusion  & 16 & 2.9 & 1.25 & 2.32\tabularnewline
\cline{3-7}
Smartphone & Apple & A9  & 16 & 4.3 & 1.05 & 4.10\tabularnewline
\cline{3-7}
 &  & A8  & 20 & 5.9 & 0.89 & 6.63\tabularnewline
\cline{2-7}
 & HiSilicon  & Kirin 960  & 16 & 5.3 & 1.10 & 4.82\tabularnewline
\cline{2-7}
 &  & Exynos 8895  & 10 & 2.9 & 1.06 & 2.74\tabularnewline
\cline{3-7}
 & Samsung & Exynos 7420  & 14 & 5.5 & 0.78 & 7.05\tabularnewline
\cline{3-7}
 &  & Exynos 5433  & 20 & 6.1 & 1.13 & 5.40\tabularnewline
\hline
\end{tabular}
\par\end{centering}
\caption{Parameters of different chips.}
\end{table}

\subsection{Heat Dissipation Analysis for Smartphones}
Considering the heat transferred from the LNAs, the total heat of the chip is calculated as ${H_{{\rm{Total}}}} = H_{{\rm{Thermal}}}^{{\rm{chip}}} + \lambda {H_{{\rm{LNA}}}}$. When ${H_{{\rm{Total}}}}$ is smaller than or equal to ${P_{{\rm{TD}}}}$, we assume that the surface temperature ${T_{{\rm{sur}}}}$ of smartphones is a constant equal to the ambient temperature ${T_{{\rm{envir}}}}$, i.e., $27\;^\circ {\rm{C}}$ or ${\rm{300}}\;{\rm{K}}$, to simplify the analysis. In this case, the heat dissipation power of smartphones is equal to ${H_{{\rm{Total}}}}$. When ${H_{{\rm{Total}}}}$ is higher than ${P_{{\rm{TD}}}}$, the extra heat $Q$ that cannot entirely be dissipated by smartphones causes a local high-temperature area on the smartphone surface. In this case, the heat dissipation power of smartphones is equal to ${P_{{\rm{TD}}}}$.

Most of the metal material used to produce the smartphone surface is 7075-T6 aluminum, which is a type of aluminum alloy and is used in the iPhone 7. The specific heat and density of 7075-T6 aluminum are $\ensuremath{\mathbb{C}=870\thinspace\rm{J/kg}\cdot{\rm K}}$ and $\rho {\rm{ = }}3000\;{\rm{kg/}}{{\rm{m}}^3}$, respectively. Moreover, the smartphone surface in this article is assumed to adopt 7075-T6 aluminum. The local high-temperature area near the chip is assumed to be a rectangle whose area $A$ is 1 ${\rm{c}}{{\rm{m}}^2}$, and the thickness of the smartphone surface $D$ is 1 ${\rm{mm}}$. Therefore, the relationship between $P_{{\rm{comp}}}^{{\rm{chip}}}$ and ${T_{{\rm{sur}}}}$ can be established by the extra heat $Q$, $\ensuremath{\ensuremath{Q=\left(P_{\rm{comp}}^{\rm{chip}}+\lambda H_{\rm{LNA}}-P_{\rm{TD}}\right)t=\mathbb{C}M\left(T_{\rm{sur}}-T_{\rm{envir}}\right)}}$, where $M$ is the mass of the surface material with volume $V = AD$ and $t$ is the stable working duration of the chip.

The extra heat $Q$ raises the temperature of the smartphone surface from ${T_{{\rm{envir}}}}$ to ${T_{{\rm{safe}}}}$ when ${H_{{\rm{Total}}}}$ is greater than ${P_{{\rm{TD}}}}$. Then, the chips in smartphones have to decrease the computation capability to reduce the heat generation. Reducing the computation capability of chips usually means that the smartphones cannot work at the original receiving rates, and the worst case is to shut off wireless communications.

\section{Maximum Receiving Rates for Smartphones}
To maintain stable computations and communications for smartphones, the value of ${H_{{\rm{Total}}}}$ must be less than or equal to ${P_{{\rm{TD}}}}$. Considering that the total proportion of ${P_{{\rm{AP}}}}$ and ${P_{{\rm{Storage}}}}$ is at least 64 \% in $P_{{\rm{comp}}}^{{\rm{chip}}}$ \cite{14Mohan, 15Ogawa}, the maximum value of $\beta $, which is the proportion of ${P_{{\rm{BP}}}}$ in $P_{{\rm{comp}}}^{{\rm{chip}}}$, is 34 \%. Moreover, the maximum receiving rate of smartphones is limited by the computation capability of the BP and the thermal design power of smartphones. Thus, the maximum receiving rate of smartphones can be expressed as ${R_{\max }} = \frac{{\beta \left( {{P_{{\rm{TD}}}} - \lambda {H_{{\rm{LNA}}}}} \right)}}{{{K_{{\rm{BP}}}}{F_0}\alpha {G_{\rm{S}}}kT\ln 2}}$.

In cellular networks, one important scheme that improves the spectral efficiency and maintains the reliability of communications is the link adaptive transmission scheme. The influence of ${R_{\max }}$ on the link adaptive transmission scheme for 5G BSs is discussed below. The most important parameter of the link adaptive transmission scheme is the channel state information (CSI). Based on the CSI, BSs can estimate the wireless channel capacity, which is the maximum downlink rate of BSs. Therefore, BSs can adjust parameters, such as the transmission power and channel coding, to achieve the maximum downlink rate of BSs.

In general, the CSI can be reflected in the signal-to-noise ratio (SNR). Considering the massive multiple-input multiple-output (MIMO) technology applied to 5G BSs, the maximum downlink rate of BSs for a smartphone ${R_{{\rm{downlink}}}}$ depends on the SNR when the bandwidth $BW$ and number of antennas at a smartphone $N_{{\rm{TRX}}}^{{\rm{phone}}}$ are fixed. Based on the limits of ${R_{\max }}$ and ${R_{{\rm{downlink}}}}$, the receiving rate of smartphones ${R_{{\rm{phone}}}}$ should take the smaller value between ${R_{\max }}$ and ${R_{{\rm{downlink}}}}$ in the link adaptive transmission scheme.

\section{Simulation Results and Discussions}
In this article, we take the average heat dissipation power of smartphones as ${P_{{\rm{TD}}}}{\rm{ = }}3\;{\rm{W}}$ \cite{15Ogawa}. The detailed simulation parameters are shown in Table II. Without loss of generality, 5-nm, 10-nm and 14-nm semiconductor technologies are used for the power analysis of the chip. Furthermore, the stable communication duration, i.e., the period that the receiving rate of smartphones can be held at a specific value, is analyzed for smartphones.

\begin{table}
\begin{centering}
\begin{tabular}{|c|c|}
\hline
Parameters & Values\tabularnewline
\hline
Number of antennas in BSs & 256\tabularnewline
\hline
Transmission power of BSs & 5 W\tabularnewline
\hline
Number of antennas in smartphones & 4\tabularnewline
\hline
Noise power spectral density & -174 dBm/Hz\tabularnewline
\hline
Carrier frequency & 3.7 GHz and 28 GHz\tabularnewline
\hline
Cell radius & 100 m\tabularnewline
\hline
Bandwidth & 20 MHz and 500 MHz\tabularnewline
\hline
Power-added efficiency of LNA & 59 \%\tabularnewline
\hline
Power of LNA & 24.3 mW\tabularnewline
\hline
Ratio of the heat transferred from the LNAs to the chip & 30 \%\tabularnewline
\hline
\end{tabular}
\par\end{centering}
\caption{Simulation parameters.}
\end{table}

\begin{figure}[!t]
\begin{center}
\includegraphics[width=6.5in]{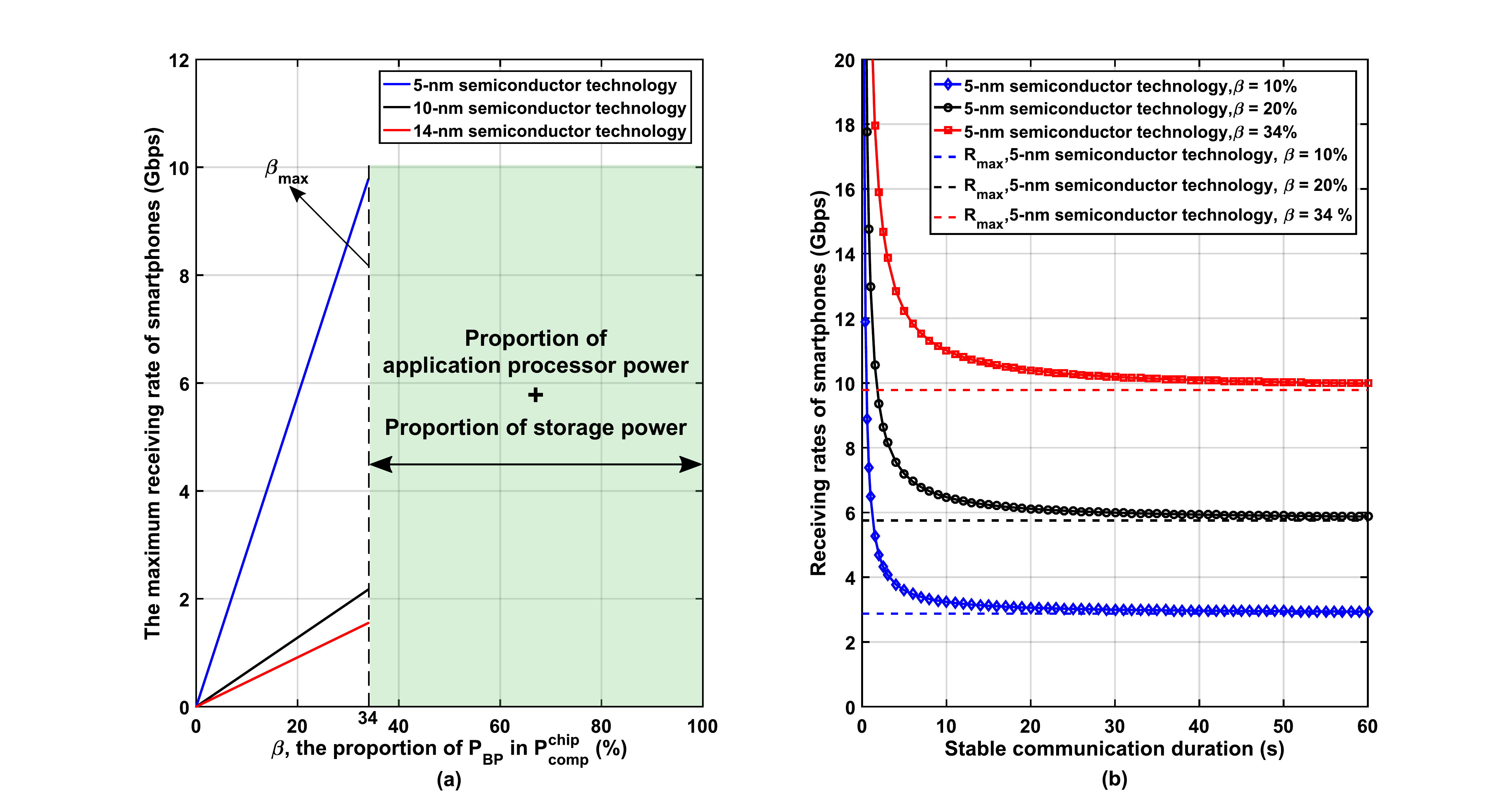}
\caption{(a) The maximum receiving rate of smartphones; (b) Receiving rates of smartphones VS. stable communication duration.}\label{Fig3}
\end{center}
\end{figure}

Fig. 3 illustrates the maximum receiving rate of smartphones as a function of $\beta $, i.e., the proportion of ${P_{{\rm{BP}}}}$ in $P_{{\rm{comp}}}^{{\rm{chip}}}$, and the stable communication duration when the rate surpasses ${R_{\max }}$. In Fig. 3a, three types of semiconductor technologies, 5-nm, 10-nm and 14-nm semiconductor technologies, have been adopted in the smartphone chip to estimate ${R_{\max }}$. Moreover, the value of $P_{{\rm{comp}}}^{{\rm{chip}}}$ is equal to ${P_{{\rm{TD}}}} - \lambda {H_{{\rm{LNA}}}}$ in Fig. 3a. When $\beta $ reaches the largest value, i.e., ${\beta _{\max }} = 34\;\% $, the values of ${R_{\max }}$ for the 5-nm, 10-nm and 14-nm semiconductor technologies are 9.74 Gbps, 2.17 Gbps and 1.55 Gbps, respectively. Based on the results in Fig. 3a, the value of ${R_{\max }}$ can be improved in the following two ways: applying the latest semiconductor technology to chips and reducing the proportions of ${P_{{\rm{AP}}}}$ and ${P_{{\rm{Storage}}}}$. When the 5-nm semiconductor technology is applied to smartphones, Fig. 3b depicts the stable communication duration when the rate surpasses ${R_{\max }}$. In Fig. 3b, when the rate is 4 Gbps, the stable communication duration is 3 seconds for ${R_{\max }}\;{\rm{ = }}\;2.9\;{\rm{Gbps}}$, whose $\beta $ value is 10 \%. Since 4 Gbps is larger than ${R_{\max }}$, whose value is 2.9 Gbps, the temperature of the smartphone surface reaches up to 45 $^\circ {\rm{C}}$ within a few seconds. In this case, smartphones must decrease the computation capability of the chip to reduce the heat generation, e.g., decrease the working frequency of the chip, to prevent low-temperature burns on the user's skin. Thus, smartphones cannot sustain the original receiving rate and may even have to shut off wireless communications. Based on the results in Fig. 3b, the stable communication duration depends on the value of ${R_{\max }}$. Furthermore, the value of ${R_{\max }}$ depends on the value of $\beta $.

\begin{figure}[!t]
\begin{center}
\includegraphics[width=2.2in]{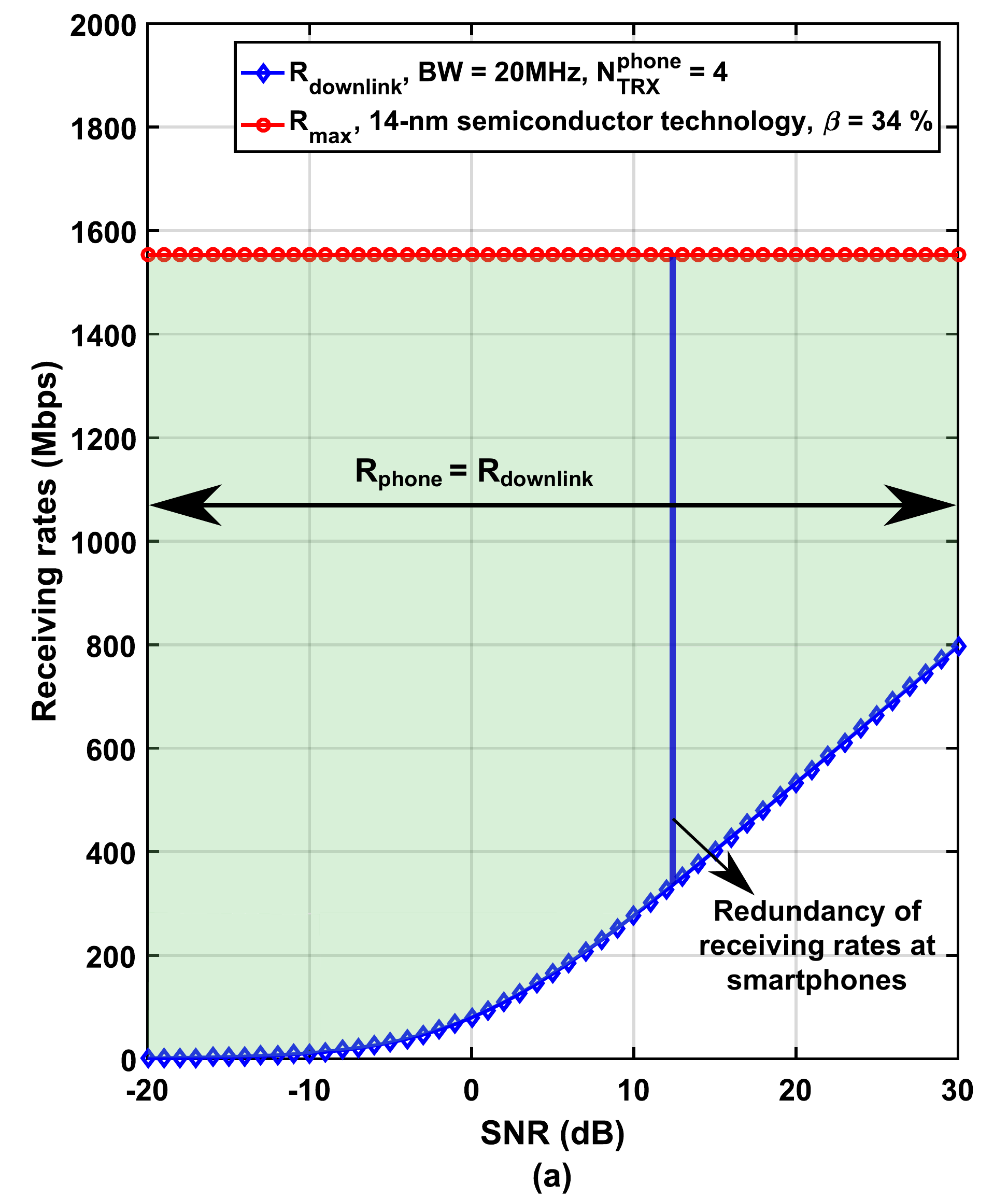}\includegraphics[width=2.2in]{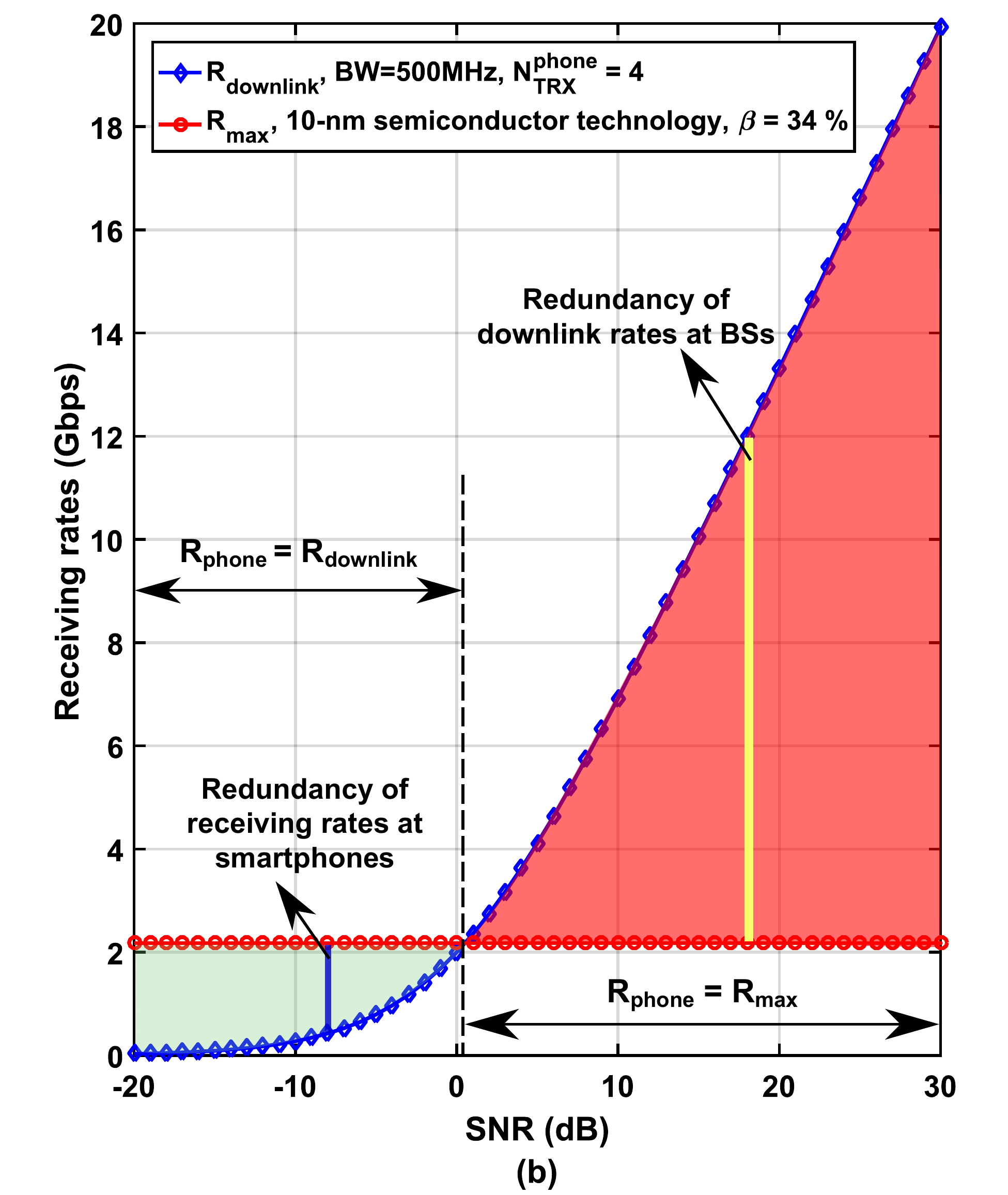}\includegraphics[width=2.2in]{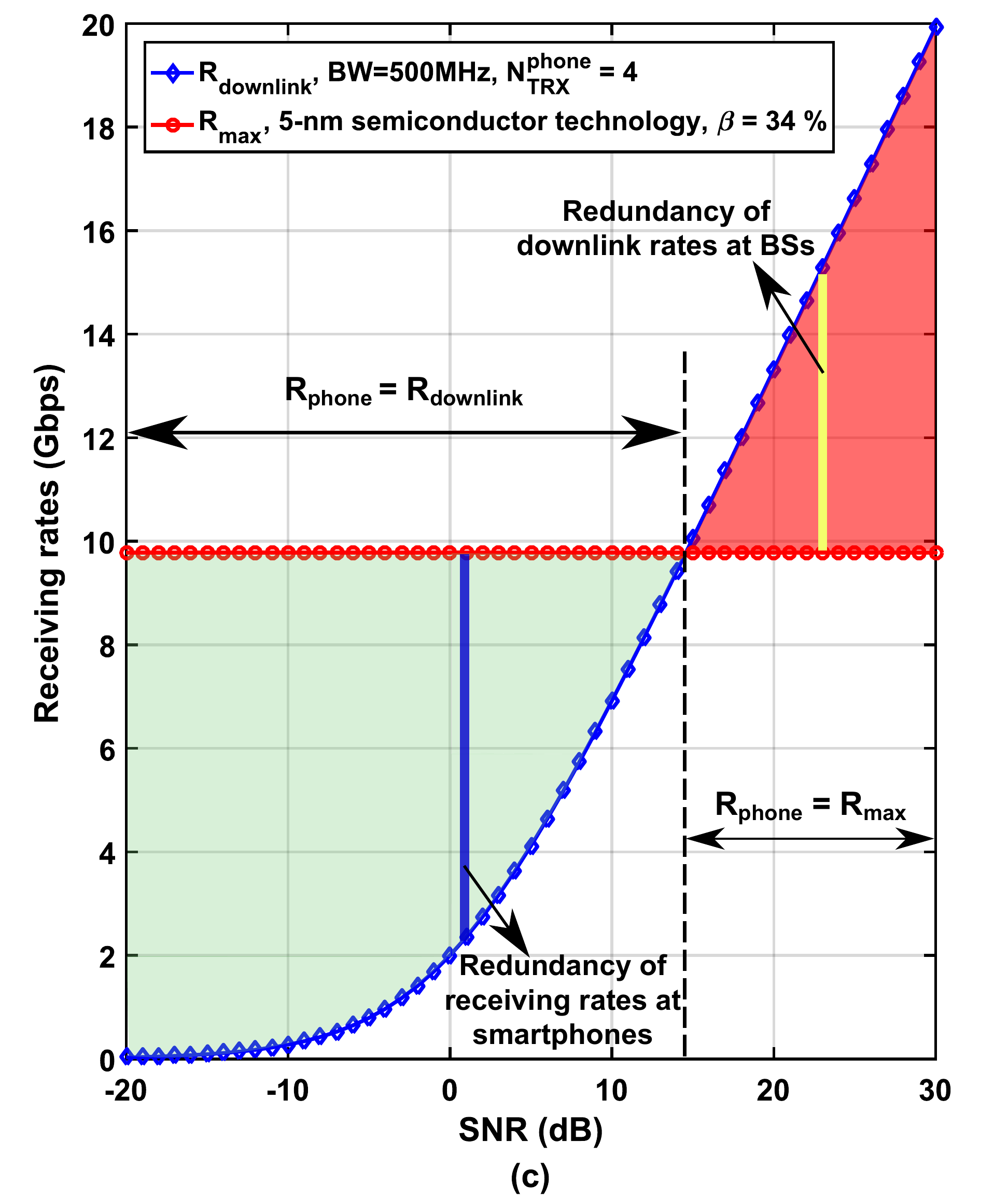}
\caption{Comparison of ${R_{\max }}$ and ${R_{{\rm{downlink}}}}$ with respect to different values of the SNR: (a) 20 MHz bandwidth and 14-nm semiconductor technology; (b) 500 MHz millimeter wave bandwidth and 10-nm semiconductor technology; and (c) 500 MHz millimeter wave bandwidth and 5-nm semiconductor technology.}\label{Fig4}
\end{center}
\end{figure}

To maintain reliable communications, the receiving rate of smartphones ${R_{{\rm{phone}}}}$ should be the smaller of ${R_{\max }}$ and ${R_{{\rm{downlink}}}}$ in the link adaptive transmission scheme. Fig. 4 illustrates the comparison of ${R_{\max }}$ and ${R_{{\rm{downlink}}}}$ with respect to different values of the SNR. In Fig. 4a, a 20 megahertz (MHz) bandwidth is used by 4G BSs, and 14-nm semiconductor technology is applied to the smartphone chip. When the SNR is less than 30 dB, the value of ${R_{\max }}$ is larger than ${R_{{\rm{downlink}}}}$. Based on the results in Fig. 4a, ${R_{{\rm{phone}}}}$ takes the value of ${R_{{\rm{downlink}}}}$ in the link adaptive transmission scheme for 4G BSs. Moreover, the redundancy of the receiving rates at smartphones in Fig. 4a, which is the gap between ${R_{\max }}$ and ${R_{{\rm{downlink}}}}$, indicates that the baseband processor of smartphones has a redundant computation capability to support a higher receiving rate. In Fig. 4b, a 500 MHz millimeter wave bandwidth is used by 5G BSs, and 10-nm semiconductor technology is applied to the smartphone chip. When the SNR is less than or equal to 0.5 dB, ${R_{{\rm{phone}}}}$ takes the value of ${R_{{\rm{downlink}}}}$. When the SNR is larger than 0.5 dB, ${R_{{\rm{phone}}}}$ takes the value of ${R_{\max }}$. In Fig. 4c, a 500 MHz millimeter wave bandwidth is provided by 5G BSs, and 5-nm semiconductor technology is assumed to be applied to the smartphone chip. When the SNR is less than or equal to 14.6 dB, ${R_{{\rm{phone}}}}$ takes the value of ${R_{{\rm{downlink}}}}$. When the SNR is larger than 14.6 dB, ${R_{{\rm{phone}}}}$ takes the value of ${R_{\max }}$. Based on the results in Fig. 4b and Fig. 4c, the value of ${R_{\max }}$ is not always larger than ${R_{{\rm{downlink}}}}$, and the receiving rate of smartphones ${R_{{\rm{phone}}}}$ should be the smaller of ${R_{\max }}$ and ${R_{{\rm{downlink}}}}$ in the link adaptive transmission scheme for 5G BSs. Furthermore, the redundancy of the downlink rates at BSs in Fig. 4b and Fig. 4c, which is the gap between ${R_{{\rm{downlink}}}}$ and ${R_{\max }}$, indicates that the baseband processor of smartphones has no extra computation capability to make the maximum receiving rate of smartphones catch up with the maximum downlink rate of BSs. A way to reduce the redundancy of downlink rates at BSs is to improve the value of ${R_{\max }}$. Based on the results in Fig. 4, the maximum receiving rate of smartphones needs to be considered in the link adaptive transmission scheme for 5G BSs.

\section{Future Challenges}
The maximum receiving rate of smartphones is not only restricted by the computation capability of the BP but also limited by the heat dissipation of smartphones. The computation capability of BP is an important challenge now that Moore's law is no longer valid and given the projected scaling limit of 5-nm gate lengths for silicon transistors. Moreover, the heat dissipation of smartphones is restricted by the safe temperature bound on the smartphone surface and the thermal design power of smartphones. In consideration of the maximum receiving rate of smartphones, two potential challenges in 5G cellular networks are presented here.

The first challenge is the impact of the maximum receiving rate of smartphones on the wireless transmission for 5G BSs, e.g., the link adaptive transmission scheme. In future studies related to the link adaptive transmission scheme for 5G, it is necessary to compare the values of the maximum receiving rate of smartphones and the maximum downlink rate of BSs. Although the maximum downlink rate of BSs can easily be estimated for 5G BSs from the CSI, the maximum receiving rate of smartphones is difficult to directly estimate. Hence, the impact of the maximum receiving rate of smartphones on the link adaptive transmission scheme for 5G BSs needs to be further investigated.

The second challenge is the trade-off between the communication and computation capabilities in 5G BSs. Based on the results in Fig. 4, redundancy of downlink rates at BSs occurs when the maximum receiving rate of smartphones is less than the maximum downlink rate of BSs. One way to reduce the redundancy of downlink rates at BSs is to apply mobile edge computing technology to offload the computation assignments of the AP and BP at smartphones. Offloading the computation assignments of the AP and BP can improve the maximum receiving rate of smartphones. Therefore, substantial computational resources will be allocated to 5G BSs. An ongoing problem is how to trade off the relationship between communication and computation capabilities in 5G BSs.

\section{Conclusions}
In order to answer the question of whether there a maximum receiving rate of smartphones to maintain stable wireless communications in 5G cellular networks, investigations on the semiconductor technology, the heat transfer and generation of the chip, and the heat dissipation of smartphones had been done. Utilizing Landauer's principle and the heat transfer theory, the maximum receiving rate of smartphones is derived for 5G communication systems. Based on the maximum receiving rate of smartphones, the potential impacts on the link adaptive transmission scheme for 5G BSs have been highlighted, e.g., the wireless communications between 5G BSs and smartphones may be shut off when the downlink rate larger than the maximum receiving rate of smartphones. To mitigate these impacts, the maximum receiving rate of smartphones should be improved by the thermal design of future 5G smartphones, e.g., using new materials or redesigning the components' structure to improve the heat conduction rate from the chip to other low-temperature components in smartphones. Additionally, mobile edge computing, one of the 5G technologies, can be applied to improve the maximum receiving rate of smartphones by offloading the computation assignments in the chips. In 5G and future 6G cellular networks, most of research is focused on the core networks and BSs. However, many potential impacts triggered by the maximum receiving rate of smartphones have not yet been investigated. How to design reasonable mobile terminals for matching with 5G and future 6G wireless communication systems is still an open issue for industries and academic researchers.

\section*{Acknowledgment}
The authors would like to acknowledge the support from the National Natural Science Foundation of China through grant 61701183, Hubei Provincial Science and Technology Department under Grant 2016AHB006, in part by EU under Grant FP7-PEOPLE-IRSES and the Project CROWN under Grant 610524, in part by the Project EXCITING under Grant 723227,  and the Fundamental Research Funds for the Central Universities through grant 2018KFYYXJJ139. This research is partially supported by the Graduates' Innovation Fund, Huazhong University of Science and Technology.

\section*{Biographies}
JING YANG [S'17] (yang\_jing@mail.hust.edu.cn) received his B.E. degrees in communication engineering from HUST in 2014. He is currently pursuing the Ph.D. degree in the School of Electronic Information and Communications at HUST. His research interests mainly include computation power of wireless communication systems, green communication, energy efficiency of wireless cellular networks and hybrid precoding.
\\

XIAOHU GE [M'09-SM'11] (xhge@mail.hust.edu.cn) is currently a full professor with the School of Electronic Information and Communications at Huazhong University of Science and Technology (HUST), China, and an adjunct professor with the Faculty of Engineering and Information Technology at the University of Technology Sydney (UTS), Australia. He received his Ph.D. degree in communication and information engineering from HUST in 2003. His research interests include green communications, vehicular communications and wireless networks. He is the director of the China International Joint Research Center of Green Communications and Networking. He serves as
associate editors for IEEE Transaction on Greening Communications and Networking, IEEE Transactions on Vehicular Technology, etc.
\\

YI ZHONG [S'12-M'15] (yzhong@mail.hust.edu.cn) is an assistant professor with School of Electronic Information and Communications, Huazhong University of Science and Technology, Wuhan, China. He received his B.S. and Ph.D. degree from University of Science and Technology of China (USTC) in 2010 and 2015 respectively. After that, he was a Postdoctoral Research Fellow with the Singapore University of Technology and Design (SUTD) in the Wireless Networks and Decision Systems (WNDS) Group.
\\

\end{document}